\documentclass[preprint,showpacs,preprintnumbers,amsmath,amssymb,superscriptaddress]{revtex4}

\include{ams}
\usepackage{amsmath,amssymb,amsthm}
\usepackage{dcolumn}
\usepackage{bm}
\usepackage{graphicx}
\usepackage{longtable}
\begin{document}
\title[a-TLS in Amorphous Organic Materials]{Anomalous Tunneling Systems in Amorphous Organic Materials}

\author{S. Sahling}
\affiliation{Institut f\"{u}r Festk\"{o}rperphysik, TU Dresden,
01062 Dresden, Germany}
\author{M.  Kol\'{a}\v{c}}
\affiliation{Charles University, Prague, Czech Republic}
\author{V.L. Katkov}\email{katkov@theor.jinr.ru}
\author{V.A. Osipov}
\affiliation{Bogoliubov Laboratory of Theoretical Physics, Joint
Institute for Nuclear Research, 141980 Dubna, Moscow region,
Russia}

\begin{abstract}
We compare the heat release data of organic glasses with that of amorphous and glass like crystalline solids. Anomalous behavior was found in all these materials, which disagrees with the standard tunneling model. We can explain the most of the experimental observations within a phenomenological model, where we assume that for a part of tunneling systems the barrier heights are strongly reduced as a consequence of the local stress produced during the cooling process.  
\end{abstract}

\pacs{72.15.Eb, 65.40.Ba, 66.70.Hk, 73.40.Gk}


\maketitle

\section{Introduction}

\begin{figure}[t]
\centering
\includegraphics[width=12.5cm,  angle=0]{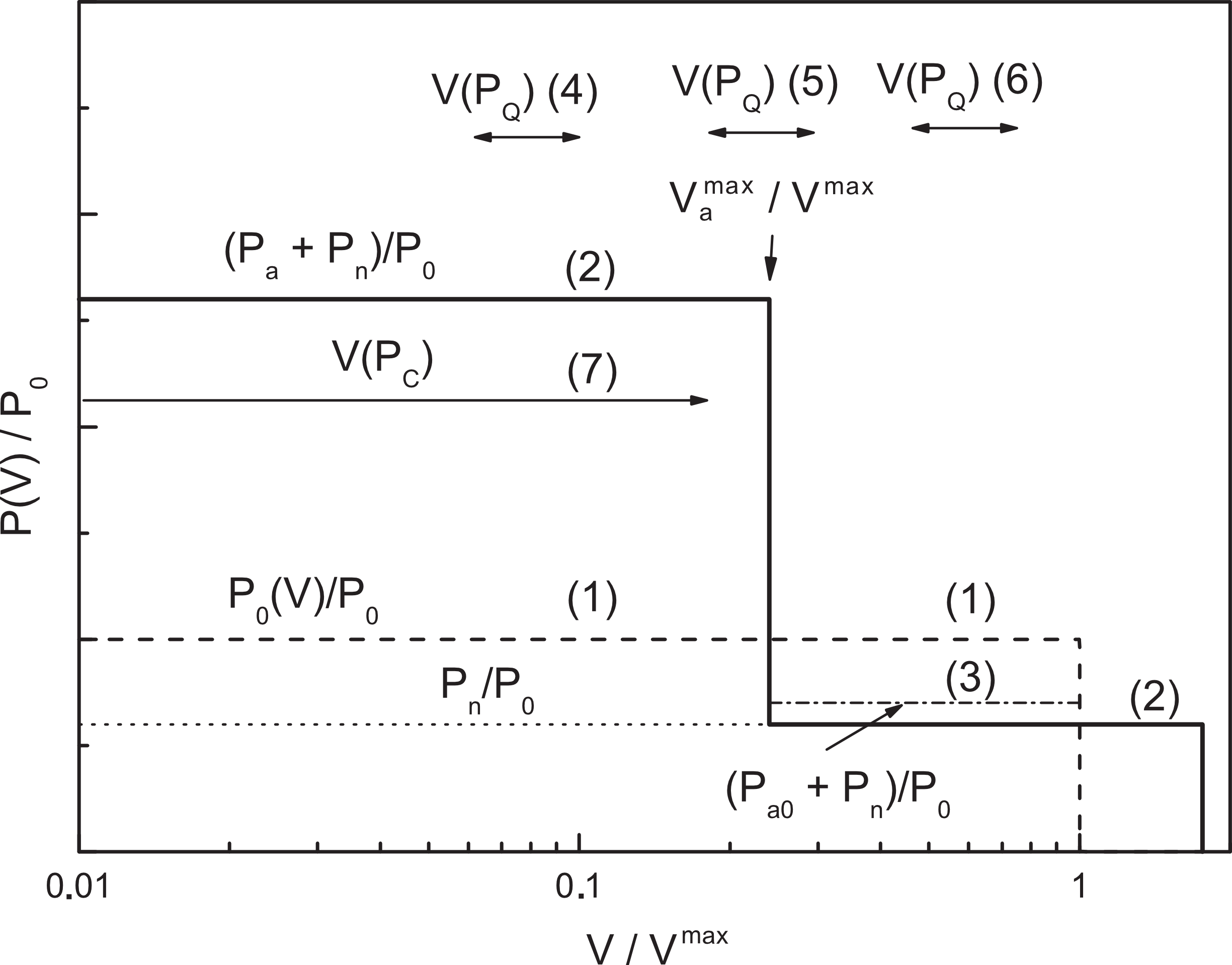}
\caption{The normalized distribution function vs the relative barrier height $V/V^{max}$: (1) corresponds to the standard tunneling model with a constant distribution up to the maximum barrier height $V^{max}$, (2) shows the distribution function according to the model of anomalous tunneling systems. The local stress reduces the barrier heights for a part of tunneling systems (anomalous tunneling systems) and their maximum barrier height is reduced to $V_a^{max}$. At $V > V_a^{max}$ the distribution function is mainly determined by normal tunnelling systems, where the local stress increases the barrier heights or the barrier heights remain unchanged (no stress). A small amount of aTLSs appears in this range too due to the distribution of the local stress (the curve 3). The arrows (4), (5) and (6) indicate the possible ranges of effective barrier heights corresponding to the time windows of  heat release measurements. The arrow (7) corresponds to a time range of the heat capacity measurement.}
\label{fig:1}
\end{figure}

Let us begin with an overview of the main results and conclusions concerning the heat capacity and heat release experiments in inorganic and glasslike crystalline materials.
It has been recently shown that barrier heights of two-level systems (TLSs) in both amorphous and glasslike crystalline solids can be drastically reduced by  frozen in local mechanical stresses caused by a rapid cooling of the sample \cite{c1}. The value of the reduction can be estimated in the framework of Eyring model \cite{c05} that is widely employed  for describing stress-induced yielding and non-linear mechanical response in polymer glasses \cite{c0}. Strong local fluctuations of the stress are expected to exist in glasses and the maximum barrier height $V_{max}$ is reduced to $V_{a}^{max}$ after the cooling of the sample with $V^{max} - V^{max}_a = \sigma^{max} V_{ac}$ where $V_{ac}$ is a so-called activation volume (a typical volume required for a molecular shear rearrangement) and  $\sigma$ is the frozen in stress caused by the thermal expansion.
As a consequence, the distribution of the barrier heights becomes much more complicated than that assumed within the standard tunneling model (STM) \cite{c2, c3}, which gives roughly a constant distribution $P(V) = P_0$. Namely, there appears a step in the distribution function at $V_a^{max}$ (see Fig.~\ref{fig:1}). Moreover, the relaxation time of the tunneling process also exhibits a step at the corresponding relaxation time $\tau_a^{max}$. 

In the one-phonon approximation, the relaxation rate of a thermally activated process is written as    
\begin{equation}\label{eq.3}
 \tau_t^{-1}  = A(E\Delta_0^2/k_B^3)\coth(E/2k_BT).     
\end{equation}
One gets  $\tau_t=\tau_a^{max}$ at $\Delta_0=\Delta_0^{min}$
where the minimum tunneling energy
\begin{equation}
\Delta_0^{min} = (2E_0/\pi)  \exp(-V_a^{max}/E_0),
\end{equation}
$E_0$ is the zero-point energy \cite{c4}, and 
\begin{equation}
 A = \frac{8 \pi^3 k_B^3}{\rho h^4}\left(\frac{\gamma_l^2}{\upsilon^5_l}+\frac{\gamma_t^2}{\upsilon^5_t}\right).
\end{equation}
TLSs with the reduced barrier heights were called anomalous (aTLSs). The results agree well with the predictions of the STM for all parameters when the experimental time scale $t$ is less than $\tau_a^{max}$. For example, in vitreous silica $\tau_a^{max}$ is found to be of the order of $100$ s, and the spectral density obtained from the heat capacity ($P_C$), thermal conductivity, ultrasound, damping (internal friction) agree quite well with the STM \cite{c5}. However, the heat release was measured at longer time up to $10^5$ s and for $t > \tau_a^{max}$ gives  a spectral density  $P_Q$  roughly $4$ times smaller than the spectral density $P_C$ deduced from the heat capacity. One can estimate the value of the step as $(P_a+P_n)/P_n = P_C/P_Q$, where a small contribution of anomalous TLSs to the heat release at $t > \tau_a^{max}$  ($P_{a0}$ in Fig.~1) has been neglected (explanation see below). We get $P_a/P_n = 3$  for vitreous silica, which means that the main part of TLSs is anomalous \cite{c5}.  It should be stressed that the most of thermodynamical parameters taken from experiments do not allow us to distinguish between the  contribution of normal and anomalous TLSs, while this is possible by the heat release measurements. 
 
The heat capacity experiment yields the spectral density  $P_C$, which  is an average value of all tunneling systems between  $\tau^{min}$ and $t$, where $\tau^{min}$ is the minimum relaxation time of the TLSs and $t$ is the time of the heat capacity measurement (typically, $t\sim 1$ s). Thus, the heat capacity always gives the spectral density for $t < \tau_a^{max}$, that is $P_C = P_a + P_n$.

The heat release is caused by TLSs with the relaxation time close to the measuring time $t$ (numerical calculations with a standard distribution function show that the heat release at the time $t$ is produced by TLSs with the relaxation time  $t/10 < \tau_t < 20 t$). Their corresponding  barrier heights are located in a quite small range $\Delta V \approx (0.1-0.2) V_m$, where $V_m$ is an average value, which determines the relaxation time $\tau_t = t$. Thus, the heat release experiments give us the spectral density for a well-defined barrier height $V_m$ and, therefore, by increasing time one can move from the left to the right side of the step in the distribution $P(V)$.  Notice that the new parameter $\tau_a^{max}$ plays an important role in our model. Namely, in accordance with Fig.~\ref{fig:1} we get different results for both the heat release and its correlation to the heat capacity in three different regions: 
\begin{enumerate}
\item $V_m < V_a^{max}$ and $t < \tau_a^{max}$
\item  $V_m = V_a^{max}$ and $t = \tau_a^{max}$
\item  $V_m > V_a^{max}$ and $t > \tau_a^{max}$
\end{enumerate}
In the first region we reproduce the results of the STM with a new spectral density $P = P_a + P_n = P_C = P_Q$. Explicitly,
\begin{equation}\label{CP}
 C_p =  (\pi^2k_B^2/12)P_C T \ln(4t/\tau^{min}),  
\end{equation}
where $\tau^{min} = \tau_t(\Delta_0=E)$, and 
\begin{equation}\label{eq.8}
dQ/dt \equiv \dot{Q} = (\pi^2k_B^2/24) P_Q V_s (T_1^2 - T_0^2) t^{-1},                      
\end{equation}
with $V_s$ being the volume of the sample, $T_1$ the starting, and $T_0$ the final temperature. At low temperatures  ($T_0$, $T_1 < T_{c0}$), for the heat release we found no difference from the STM. Here $T_{c0}$ is the crossover temperature, where the relaxation rate of the tunneling $\tau_t^{-1}$ is equal to the relaxation rate of the thermally activated process. 
Above the crossover temperature the relaxation time of normal TLSs decreases rapidly with increasing temperature due to the Arrhenius law. As a consequence, a part of TLSs relaxes during the cooling process and does not contribute to the heat release. In this case, the heat release deviates from $T_1^2 - T_0^2$ behavior. At high enough temperatures $T_1 > T^*$ ($T^* $ is the freezing temperature), all TLSs with the energy  $E > 2.4 k_B T^*$ reach the state of equilibrium during the cooling process, and the heat release saturates. Thus a maximum value of the heat release will be observed at $T_1\approx T^*$  in Eq. (\ref{eq.8}). $T^*$  depends weekly on the cooling rate $R^*$ at this temperature and is directly proportional to the barrier height $V_m$ of the TLSs causing the heat release at given time after the cooling to a temperature  $T_0 < T_{c0}$.
Notice that  the freezing temperature is close to the crossover temperature $T_n^* \approx 1.2 T_{c0}$ for normal TLSs (see \cite{c6}). However, the barrier height of anomalous TLSs increases with increasing temperature and their freezing temperature $T^*_a$  can be essentially higher than $T_n^*$. In this case, a giant heat release will be observed since the heat release saturates at very high temperatures  $T_1 = T^*_a$ .

\begin{figure}[t]
\centering
\includegraphics[width=12.5cm,  angle=0]{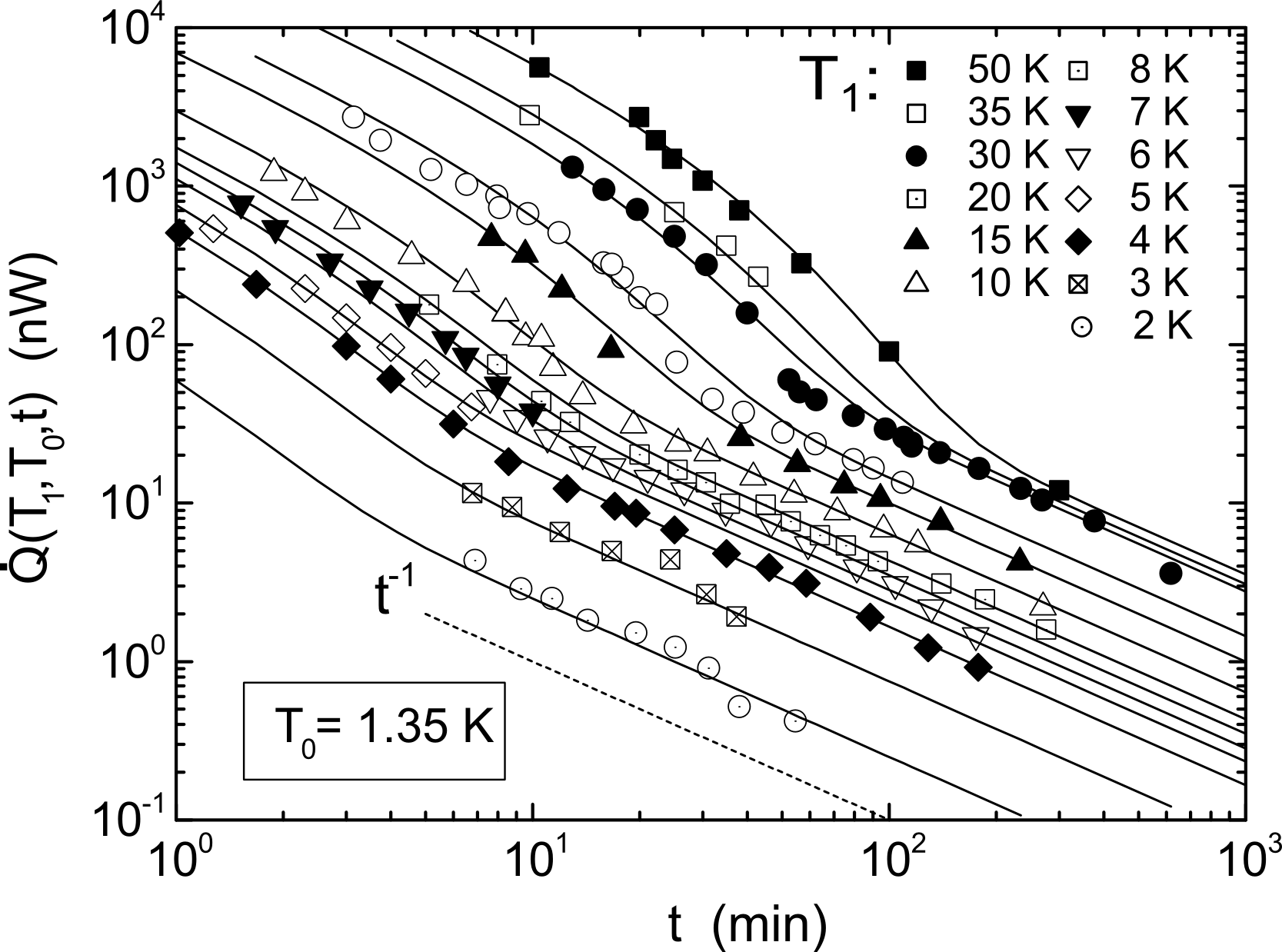}
\caption{{The heat release in 14.5 cm$^3$
(ZrO$_2$)$_{0.89}$(CaO)$_{0.11}$ after cooling from different
initial temperatures  $T_1$  to the final phonon temperature
$T_0$ as a function of time ($t = 0$ at the beginning of cooling)
\cite{c7}. The curves are calculated with Eq.~(\ref{eq.13}),
which corresponds to the distribution $P(\tau)$ of the standard
tunneling model with a step at $\tau_a^{max}$ caused by the
cut-off in the distribution of anomalous TLSs. The fit parameter
$\tau_a^{max}$ is given in Fig.~\ref{fig:3}}}
\label{fig:2}
\end{figure}

\begin{figure}[t]
\centering
\includegraphics[width=12.5cm,  angle=0]{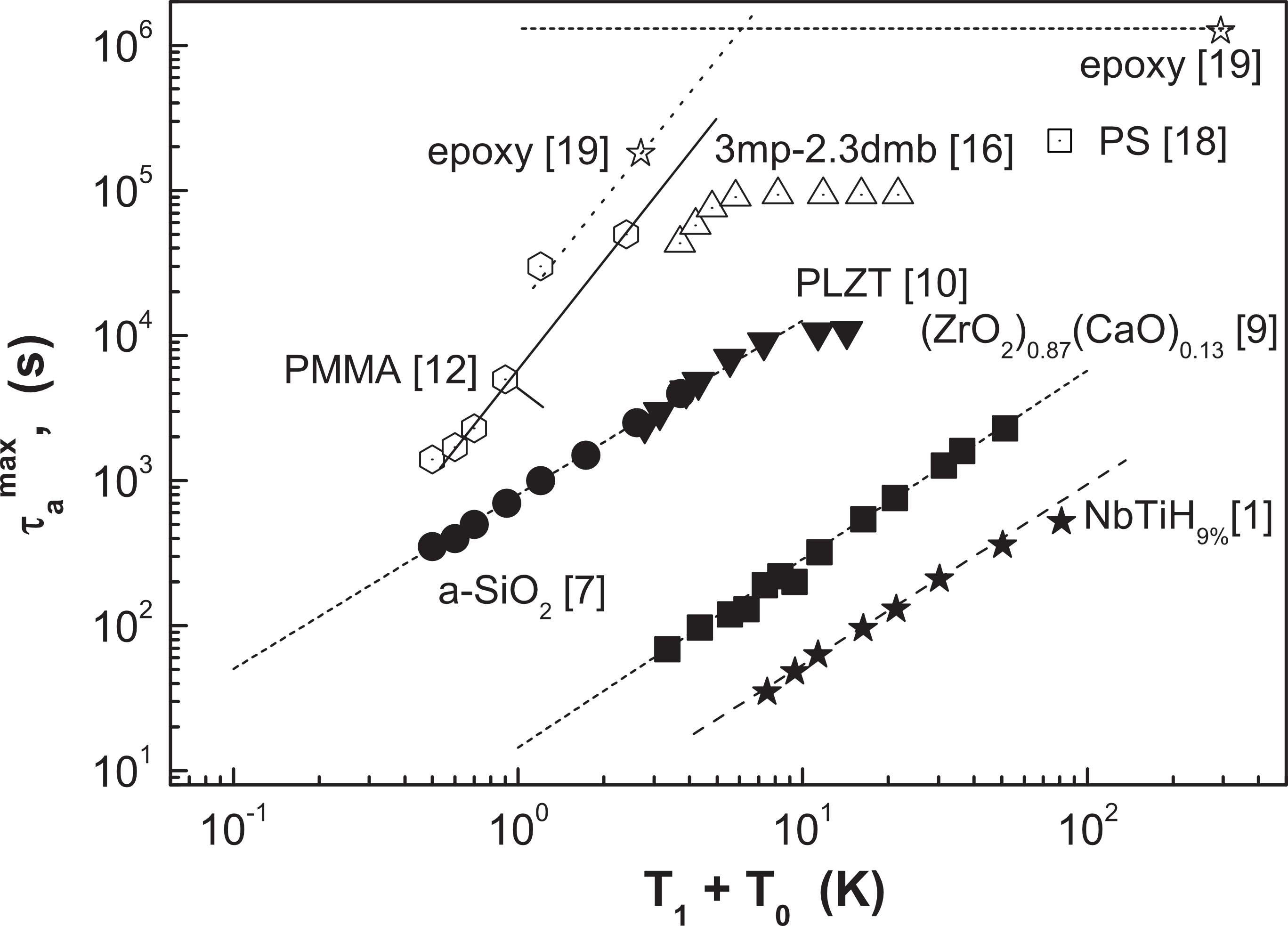}
\caption{The fit parameter  $\tau_a^{max}$ (used to fit the curves of the heat release data as a function of time for different materials (see Figs. 2,5,7-10)) as a function of $T_1 + T_0$, which is close to an average energy of relaxing tunneling systems.}
\label{fig:3}
\end{figure}

In the second region, the heat release relaxes very fast and can be described by
the relation
\begin{equation}\label{eq.13}
\dot{Q} =  [Q_l + (Q_s - Q_l) \exp(-t/\tau_a^{max})]t^{-1},  
\end{equation}
where $Q_l$ and $Q_s$ are proportional to $P_{a0} + P_n$ and $P_a + P_n$, respectively. In the heat release one can directly see the step in the distribution function  and determine the maximum relaxation time of anomalous TLSs as a function of an average energy  $E_{av}/k_B = 2.4(T_1 + T_0)/2$. This rare case was observed in ZrCaO \cite{c7}. Fig.~\ref{fig:2} shows the heat release as a function of time after cooling from different $T_1$ to $T_0 = 1.3$ K. 
Notice that Eq.~(\ref{eq.13}) describes the time dependence very well. Fig.~\ref{fig:3} shows the energy dependence of $\tau_a^{max}$, where we point out another interesting fact: the maximum energy is not proportional to $1/E_{av}$, as expected from the STM (see Eq.~(\ref{eq.3})) but roughly proportional to  $E_{av} \approx T_1 +T_0$. 

In the region 3 there is a remarkable discrepancy between  $P_C$ and $P_Q$ since
\begin{equation}
\frac{P_C}{P_Q} = \frac{P_a+ P_n}{P_{a0}+ P_n}  > 1.    
\end{equation}
In this case, the heat release is mainly determined by the normal TLSs, and the maximum heat release is much smaller than the value measured at $t < \tau_a^{max}$. Nevertheless, there is also a small contribution of anomalous TLSs  ($P_{a0}$ in Fig.~\ref{fig:1}). It was suggested that this small contribution to the heat release at $t > \tau_a^{max}$ could be caused by some distribution of the local stresses in the sample after a rapid cooling. Within this scenario, the maximum local stress will reduce  $V^{max}$ to $V_a^{max}$ after the cooling. For smaller stress, the corresponding $V$ will take the values between  $V_a^{max}$ and $V^{max}$ at the relaxation time  $t > \tau_a^{max}$, and a corresponding  freezing temperature $T^*$ varies between $T^*_n < T^* <T^*_a$. Thus, the distribution of the local stresses leads to a contribution of anomalous TLSs at $t > \tau_a^{max}$. Such a contribution was observed in all materials with  $P_C > P_Q$. Notice that $P_{a0}$ is small in comparison to  $P_a$ - for example,  $P_{a0}/P_a = 0.05$ for vitrous silica \cite{c5}. Nevertheless, for high initial temperatures this fraction of anomalous TLSs gives the dominant contribution to the heat release (roughly 2 times larger than the contribution of normal TLSs) since they have also a much higher $T^*_{a0}$ in comparison to $T^*_{n}$  (see Fig.~4). 
\begin{figure}[t]
\centering
\includegraphics[width=12.5cm,  angle=0]{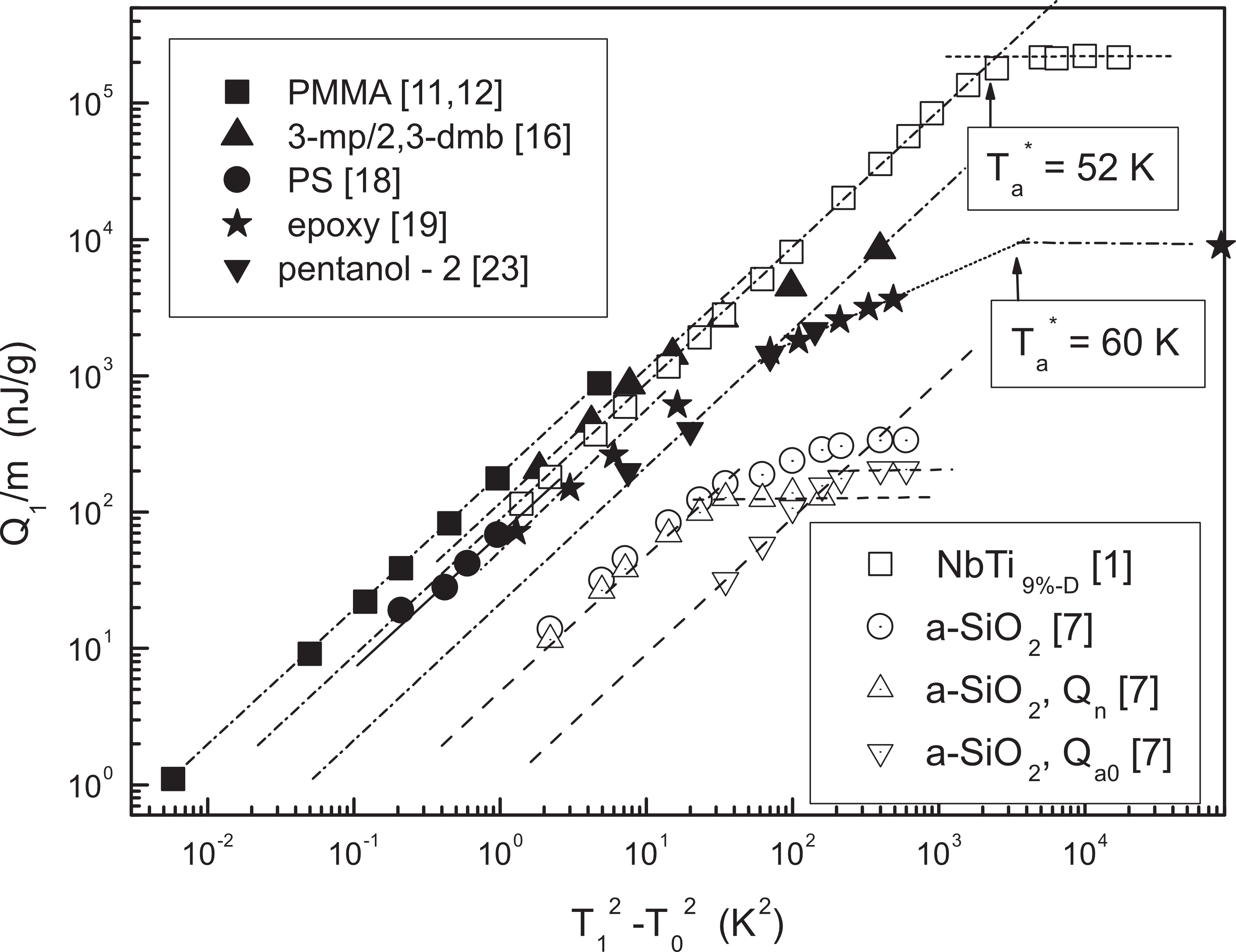}
\caption{The fit parameters  $Q_1 = t \dot{Q}$ as a function of  $T_1^2-T_0^2$ for different materials (see Figs. 2,5, 7-10). A characteristic feature of anomalous TLSs is a high freezing temperature $T^*$.}
\label{fig:4}
\end{figure}
Measuring the $T_1$-dependence of the heat release one can separate the contribution of  normal and anomalous TLSs. Fig.~4 shows the parameter  $Q_1 = t\dot{Q}$ as a function of $T_1^2 -T_0^2$. The contribution of normal TLSs saturates for a-SiO$_2$  at $T^*_n = 5.5$ K while for anomalous TLSs ($P_{a0}$) it saturates above $14.1$ K. Values of these two contributions are proportional to  $P_n$  and  $P_{a0}$, correspondingly.
$\tau_a^{max}$ depends on the values of local stresses and  $V^{max}$ as well as  on the mass of tunneling entity.  The last possibility was demonstrated for TLSs in NbTi caused by H or D in \cite{c1}. 

The heat release and heat capacity data of all investigated inorganic amorphous and glasslike crystalline solids complies with one of these three cases (see Table \ref{tab1}). A transition between cases one and three is rarely observed since changing measuring time in the heat release experiments by a factor of ten modifies the corresponding barrier height by a few percents only. All three cases were observed in ZrO$_2$CaO \cite{c7} (see Fig.~\ref{fig:2}), PLZT \cite{c8} and also in a-SiO$_2$ \cite{c5} and NbTi-H \cite{c1}. 

It should be noted that the heat release in amorphous organic materials has not yet been analysed in detail. At the same time, the experimental studies of organic glasses show a rather specific behavior of the heat release. In this paper, we make a comparison of the heat capacity and heat release data between organic glasses, inorganic glasses, and glasslike crystalline materials. Our analysis clearly indicates the existence of anomalous TLSs in all these materials; their possible origin is the local stresses during the cooling process.

\section{Heat capacity and heat release of amorphous organic materials}

Our purpose is to analyse the relevant heat release and heat capacity data for  organic amorphous materials and show that anomalous TLSs should also exist in these materials. If this is the case, the distribution function must be similar to that in inorganic glasses with a step at $V_a^{max}$. There is, however, an important difference: in organic materials the heat release decays as $t^{-a}$ with $a<1$  in contrast to both the predictions of STM and the $t^{-1}$ behavior observed in most of inorganic glasses. As a possible explanation one can assume an additional strong increase of $P(\tau)$ near $\tau_a^{max}$. Let us consider an extreme case of a $\delta$-like growth.  In this case, the heat release should be written as
\begin{equation}\label{eq.17}
 \dot{Q}   = Q_n/t + \dot{Q}_a      
\end{equation}
with 
\begin{equation}\label{eq.18}
\dot{Q}_a = \left[Q_a/t+ P_{ax}\right] \exp(-t/\tau_a^{max}),
\end{equation}
where $P_{ax}$  is a time independent contribution to the heat release. 
Otherwise, we would expect to find for  $P_Q$ a value larger or close to $P_C$ for $t<\tau_a^{max}$, and a fast relaxation of the heat release at $t \approx  \tau_a^{max}$.

\subsection {Heat capacity and heat release of PMMA}

\begin{figure}[t]
\centering
\includegraphics[width=12.5cm,  angle=0]{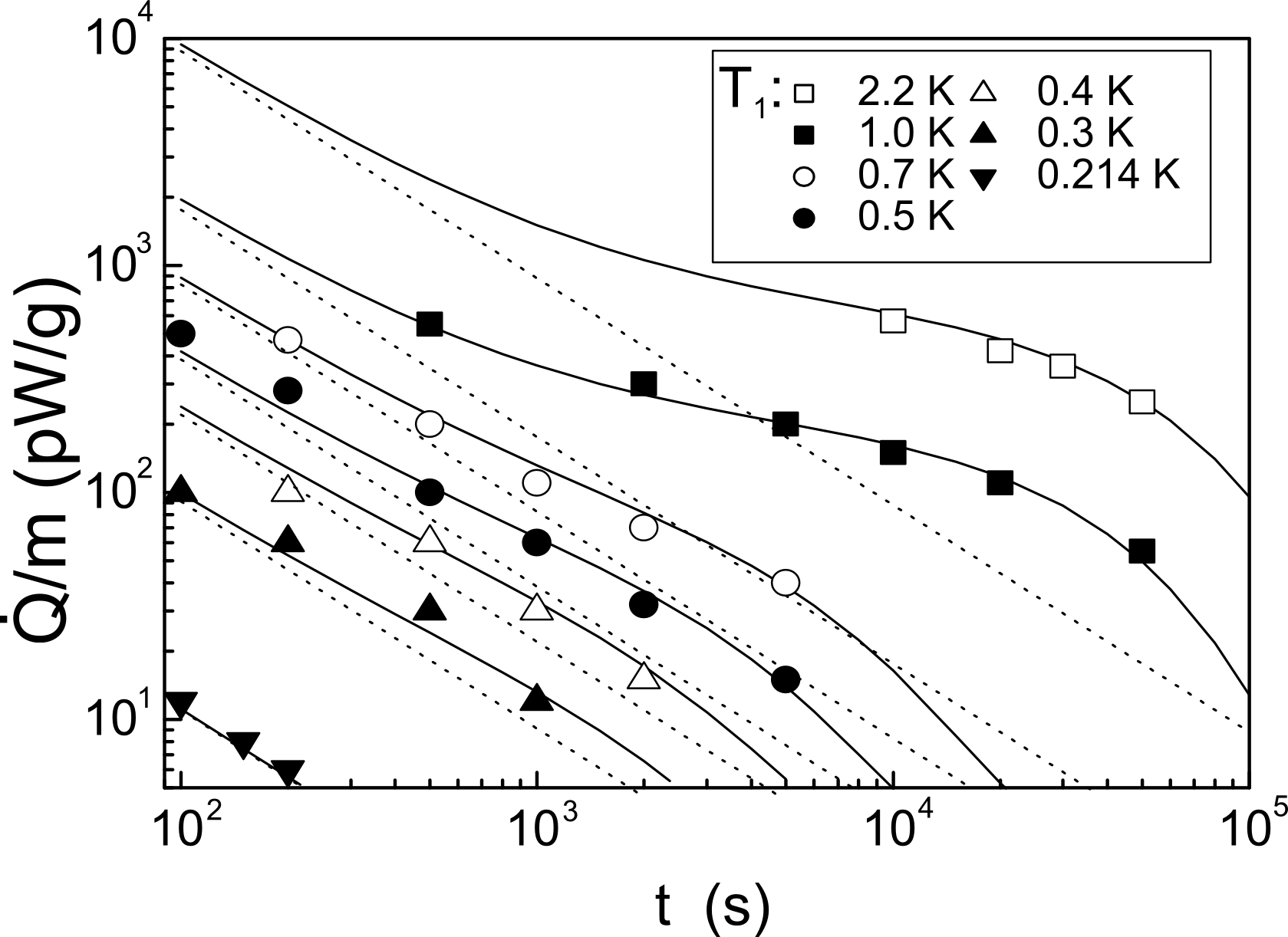}
\caption{The heat release $\dot{Q}/m$ as a function of time for PMMA. The experimental data are taken from \cite{c9,c10}. The curves correspond to Eq. (\ref{eq.19}). Dashed lines show the first term in Eq. (\ref{eq.19}). The fit parameters  $Q_1$ and $\tau_a^{max}$ are given in Figs. 3 and 4.}
\label{fig:6}
\end{figure}

\begin{figure}[t]
\centering
\includegraphics[width=12.5cm,  angle=0]{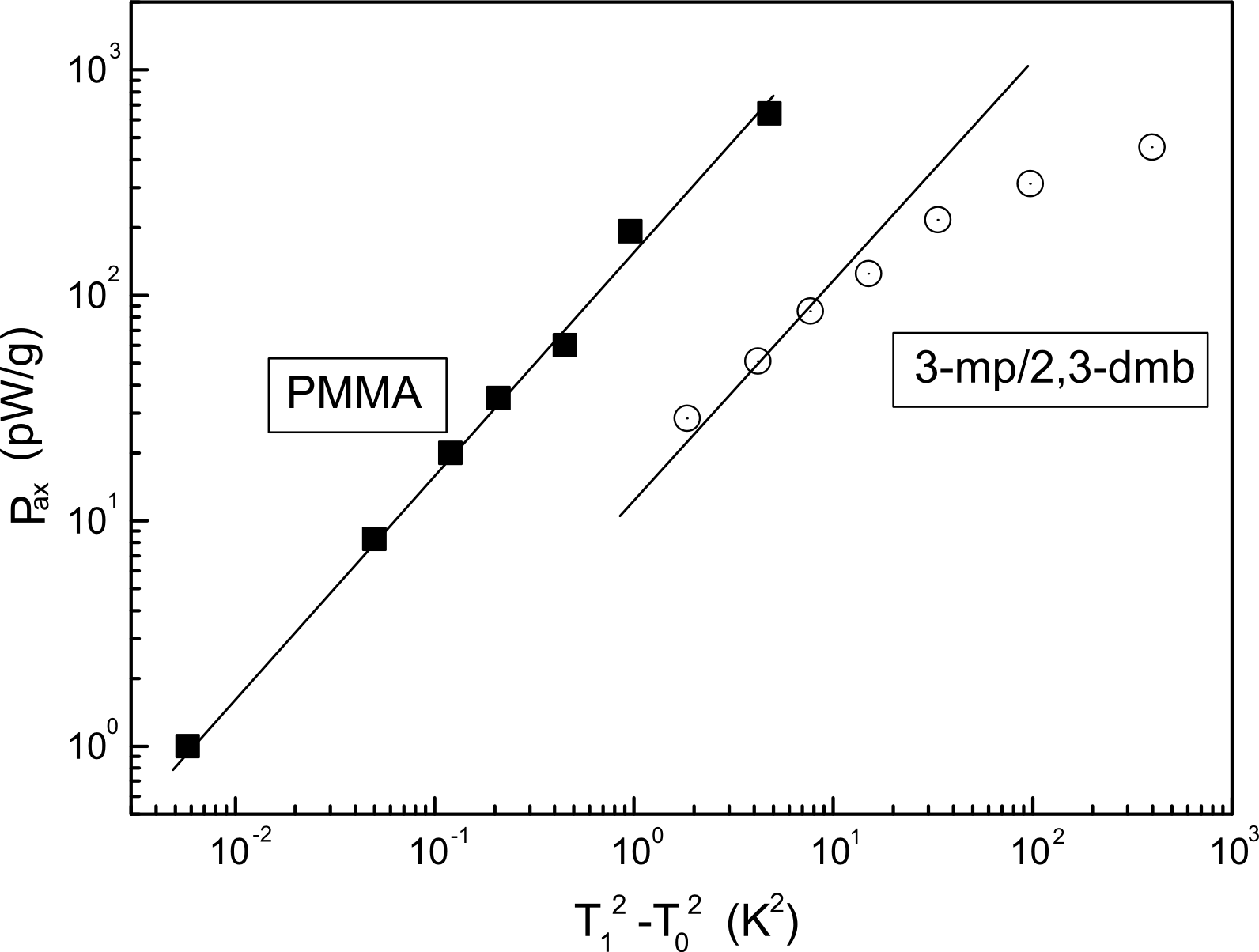}
\caption{The fit parameters  $P_{ax}$ as a function of  $T_1^2-T_0^2$ used for the calculation of the heat release data of PMMA and 3-mp/2,3-dmb with Eq. (\ref{eq.19}).}
\label{fig:7}
\end{figure}

One of the first long time heat release experiments were performed by Zimmermann and Weber with a-SiO$_2$ and PMMA \cite{c9, c10}. In both materials the poor agreement with the STM was observed.  The results of vitreous silica were considered and discussed in detail in ref. \cite{c5}. The results of PMMA are shown in Fig.~\ref{fig:6}.  For low $T_1$, the heat release is roughly proportional to $t^{-1}$ as predicted by the STM. However, a markedly different time dependence was observed for higher $T_1$.
A much better agreement we get with a fit 
\begin{equation}\label{eq.19}
\dot{Q} = (Q_1/t + P_{ax}) \exp(-t/\tau_a^{max}),
\end{equation}
where  
\begin{equation}\label{eq.20}
Q_1 = Q_a + Q_n.  
\end{equation}
The heat release of PMMA is at low temperatures $T_1$ and short time mainly determined by the term $Q_l/t$ and agrees here quite well with the STM. However, Eq.~(\ref{eq.19}) yields a much better agreement with the experimental data including that data at higher $T_1$ and long time. As is seen, that Eq.~(\ref{eq.19}) differs from  Eqs. (\ref{eq.17}) and (\ref{eq.18}): the exponent cuts down also $Q_n$, and therefore this fit fails at  $t \gg \tau_a^{max}$. In order to separate terms $Q_a$ and $Q_n$ we need the heat release data either for higher temperatures $T_1$ or for $t \gg \tau_a^{max}$. Unfortunately, they are not yet available for PMMA. As we see in Fig.~\ref{fig:4}, $T_1^2 - T_0^2$ dependence of $Q_1$ agrees with the STM. Thus, we can estimate the distribution parameter by using the STM. The result is $P_Q = 28\times 10^{44} $ Jm$^3$ at $\rho = 1.19$ g/cm$^3$. Notice that this parameter really markedly exceeds the value deduced from the heat capacity $P_C  = 8.4 \times 10^{44}$  Jm$^3$ \cite{c11}. This finding indicates that the heat release is strongly influenced by the new term $P_{ax}$ which 
is also strongly proportional to $T_1^2 - T_0^2$  (see Fig.~\ref{fig:7}). This means that all contributions to the heat release have the same or similar distribution function of the energy  ($P(E) = const$).  

Notice that the increase of the distribution function with $V$ could explain the deviation from $t^{-1}$ behavior of the heat release. However, in this case we would obtain a deviation from $T_1^2 - T_0^2$ behavior as well. In particular, the model with TLS-TLS interaction taken into account gives 
 $P\sim 1/\Delta_0^2$ \cite{Burin}. This function was successfully used to fit the data of spectral diffusion in PMMA \cite{Maier}. However, using this distribution function in our case we obiain $\dot{Q}\sim T_1^{5/2} - T_0^{5/2}$, which is inconsistent with the heat release experimental data (see Fig. \ref{fig:4}).


The contribution of the anomalous TLSs and normal TLSs cannot be separated because this requires the heat release experiments with higher $T_1$ which were not performed up to now.

\subsection{Heat release of 3-MP/2,3-DMB}

\begin{figure}[t]
\centering
\includegraphics[width=12.5cm,  angle=0]{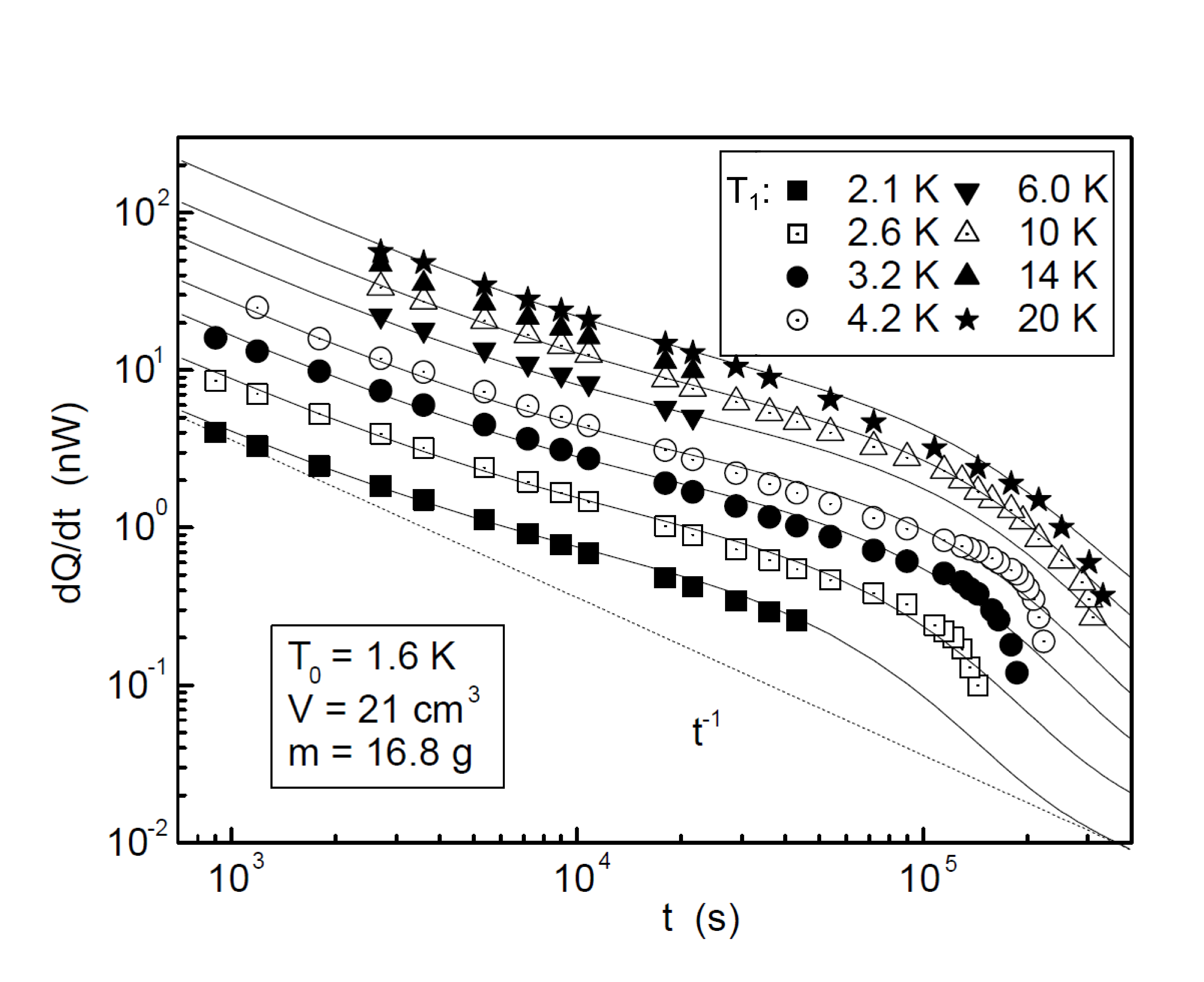}
\caption{The heat release of   21 cm$^3$ 3-mp/2,3-dmb as a function of time after cooling from different temperatures  $T_1$ (see the insert) to $1.6$ K \cite{c12}. The relaxation becomes very rapid for a long time. The curves are calculated with according to  Eq.~(\ref{eq.19}). The fit parameters  $P_{ax}$, $Q_1$ and $\tau_a^{max}$ are given in Figs. 3, 4 and 6.}
\label{fig:8}
\end{figure}

Heat release measurements were performed with 3-methylpentane/2,3-dimethylbutan at much higher starting temperature than for PMMA (up to 20 K)  \cite{c12}. The heat release is found to be roughly proportional to $t^{-0.7}$ and relaxes much faster for a very long time (see Fig.~\ref{fig:8}). Notice that even a sharp cut of the TLSs distribution function could not provoke such strong dying in $\dot{Q}$. The full lines present the fit curves according to Eq.(\ref{eq.19}). The corresponding parameters  $Q_1$  and  $P_{ax}$  are shown in Figs. \ref{fig:4} and \ref{fig:7} as a function of $T_1^2- T_0^2$.  $Q_1$ is evidently proportional to $T_1^2- T_0^2$  for  $T_1 < 3$K and one can estimate $P_Q = 16.7 \times 10^{44}$ Jm$^3$.   At higher $T_1$, the results deviate from the linear dependence. However, no saturation of the heat release as a function of $T_1$ is observed up to $20$ K, which is a typical behavior of anomalous TLSs. This is a clear indication that  $t < \tau_a^{max}$. At long time we observe a faster relaxation due to $\tau_a^{max}$. The additional contribution $P_{ax}$  determines the heat release far a long time. The energy dependence is similar to that for  $Q_1$. The parameter $\tau_a^{max}$ is close to the corresponding values of PMMA with a similar energy dependence for low $T_1$ (see Fig.~\ref{fig:3}). 

\begin{figure}[t]
\centering
\includegraphics[width=12.5cm,  angle=0]{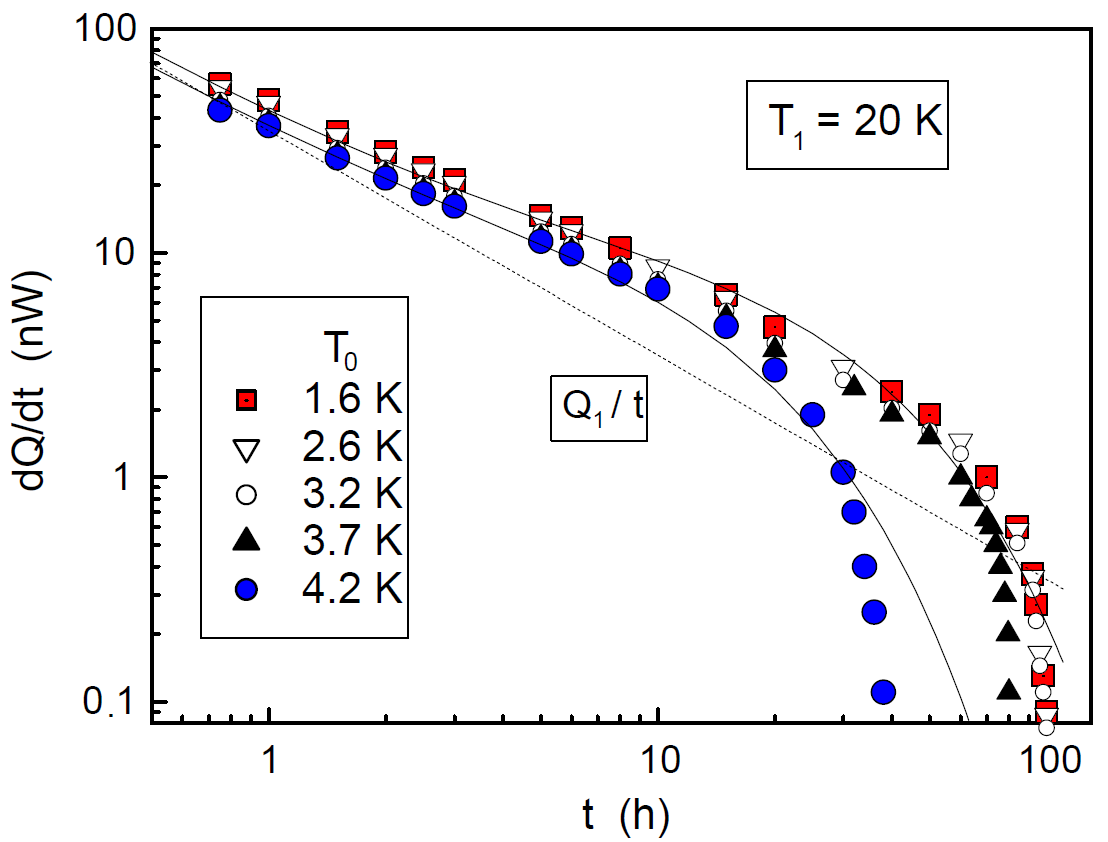}
\caption{The heat release of   21 cm$^3$  3-mp/2,3-dmb as a function of time after cooling from $20$ K to different temperatures  $T_0$ (see the insert)  \cite{c12}. The curves correspond to Eq. (\ref{eq.19}).}
\label{fig:9}
\end{figure}

Interesting results were obtained for this material by measuring the heat release as a function of the final temperature  $T_0$ (see Fig.~\ref{fig:9}). The variation in $T_0$ from 1.6 K to 3.2 K does not influence the parameter $\tau_a^{max}$. However, a strong reduction of $\tau_a^{max}$ is observed for $T_0 > 3.2$ K. This means that above $3.2$ K the thermal activation process dominates, and we expect a saturation of the heat release as a function of $T_1$ around $4$ K provided that the barrier heights are constant. In fact, $Q_1$ is no more proportional to $T_1^2- T_0^2$ for  $T_1 > 3$ K (the heat release caused by normal TLSs saturates). However, no saturation is observed up to $20$ K which is a strong indication that the further increase of the heat release at $T_1 > 4$ K  is caused by anomalous TLSs.

\subsection{Heat capacity and heat release of PS}

\begin{figure}[t]
\centering
\includegraphics[width=12.5cm,  angle=0]{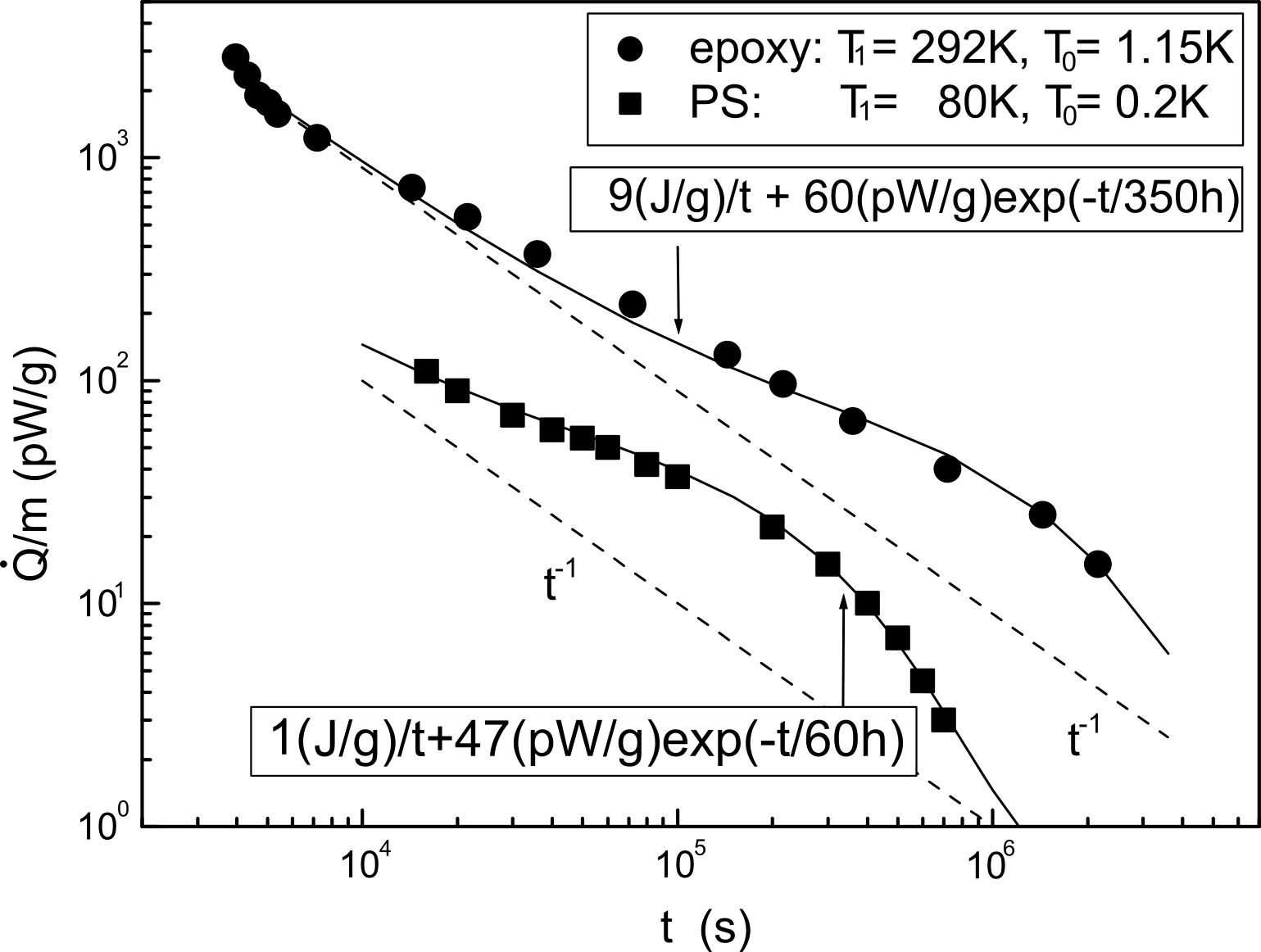}
\caption{The heat release of epoxy resin \cite{c14}  and PS \cite{Nittke} as a function of time after cooling from a very high temperature  $T_1$ to $T_0$ (for temperatures see the  insert). The curves correspond to (\ref{eq.19}).}
\label{fig:10}
\end{figure}

Heat release measurements in polystyrene were performed by Nittke et. al \cite{Nittke}.  In the range of $0.7$ h $< t <$ $6$ h at $T_0 = 0.2$ K with  $0.5$ K $< T_1 <$ $1$K measurments show that the heat release and $Q_1$ are proportional to $t^{-1}$ and $T_1^2- T_0^2$, respectively (see Fig.~17 in \cite{Nittke}). Resulting $P_Q = 9\times 10^{38}$ J/g, is in a good agreement with $P_C = 7.5\times 10^{38}$ J/g deduced from the heat capacity \cite{Nittke, c13}. All results are in perfect agreement with the STM and no contributions of anomalous TLSs was observed. Notice, however, that the heat release was measured for a quite short time and low $T_1$ only. The contribution of anomalous TLSs would be much larger for high $T_1$  and long time.
Indeed, the heat release measured after cooling from $80$ K to $0.3$ K in a wide time range shows the typical time dependence caused by the anomalous TLSs (see Fig.~\ref{fig:10}) with  $\tau_a^{max} = 60$ h.

\subsection{Heat capacity and heat release of epoxy resin}

The heat release of epoxy resin was measured after cooling from $292$ K to $1.15$ K during $25$ days \cite{c14}. It was found to be proportional to  $t^{-0.76}$ (see Fig. \ref{fig:10}). A similar time dependence was found for epoxy glue Stycast \cite{c15}. Eq.~(\ref{eq.17}) gives a good fit for epoxy with a very long $\tau_a^{max} = 350$ h. At lower $T_1$ the power law in the time dependence does not change: the relation between  $Q_1$  and  $P_{ax}$ is the same for different $T_1$. This holds only when both contributions are caused by anomalous TLSs. $Q_1$ deviates from the STM at $T_1 > 3$K  (the saturation of the heat release caused by the normal TLSs) but the heat release is still increasing up to the highest temperature $T_1 = 22$ K. This is again a clear indication that the heat release is produced by anomalous  TLSs.
\begin{figure}[t]
\centering
\includegraphics[width=12.5cm,  angle=0]{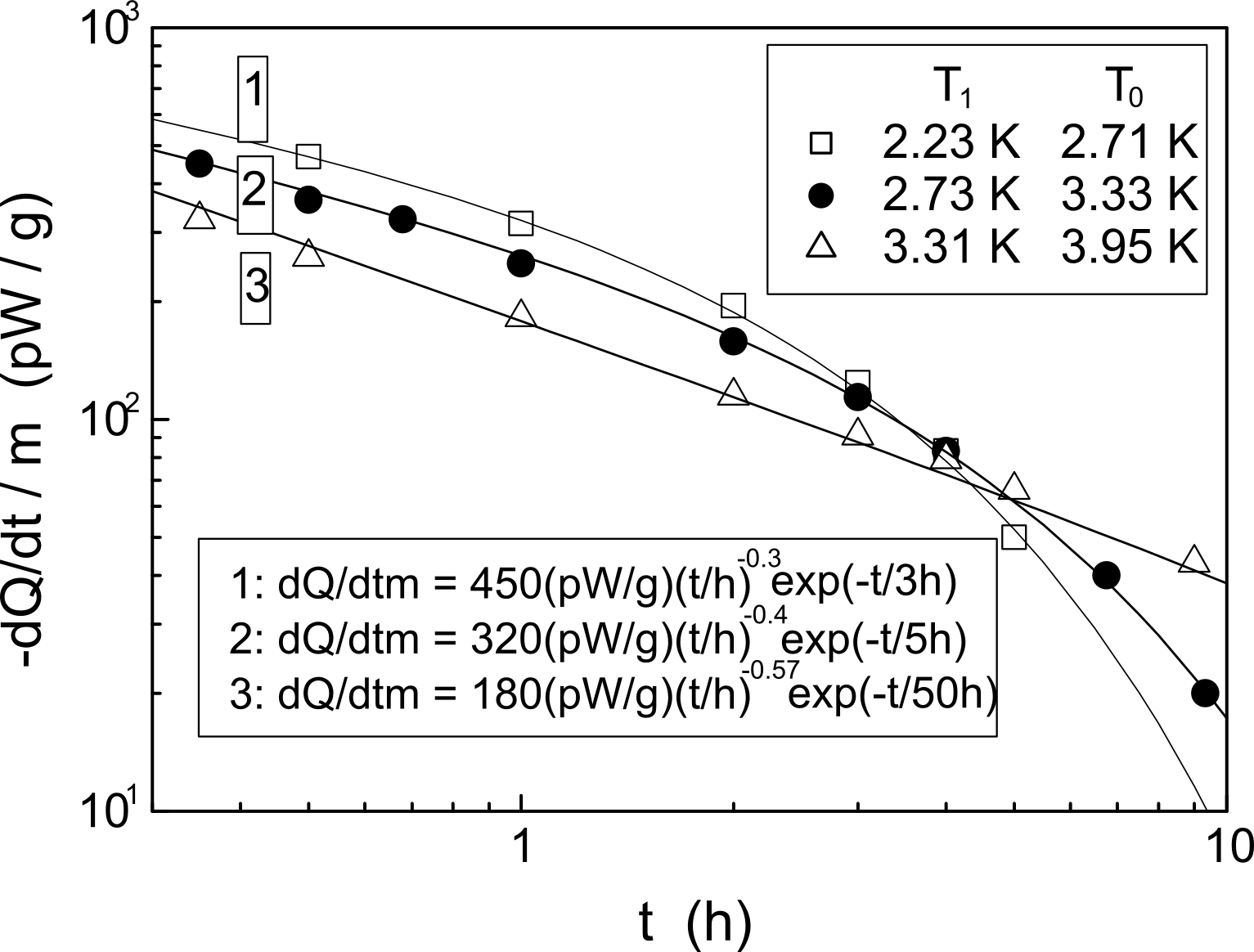}
\caption{The heat absorbed in epoxy resin after rapid heating from $T_1$  to  $T_0$ (temperatures are given in the insert) \cite{c14}. The curves correspond to Eq. (\ref{eq.21}).}
\label{fig:11}
\end{figure}

The step in the distribution function was found for higher $T_0$ in the thermally activated range \cite{c14} (see Fig.~\ref{fig:11}). Here the thermal absorption was measured after heating the sample from $T_1$ to $T_0$  (i.e.  $T_1 < T_0$). Since the thermal activation alters the time dependence for shorter time as well, we fit the data by a power law
\begin{equation}\label{eq.21}
\dot{Q}  = P(t_0) (t_0/t)^{-a} \exp(-t/\tau_a^{max}),
\end{equation}
where  $P(t_0)$ is the heat release at  $t = t_0$  ($t_0 \ll \tau_a^{max}$) and the fit parameter $a$ depends on $T_0$. The parameter  $\tau_a^{max}$ changes rapidly with increasing temperature $T_0$ from $3$ h  at  $3.95$ K  to $50$ h at $2.23$ K.
\begin{figure}[t]
\centering
\includegraphics[width=12.5cm,  angle=0]{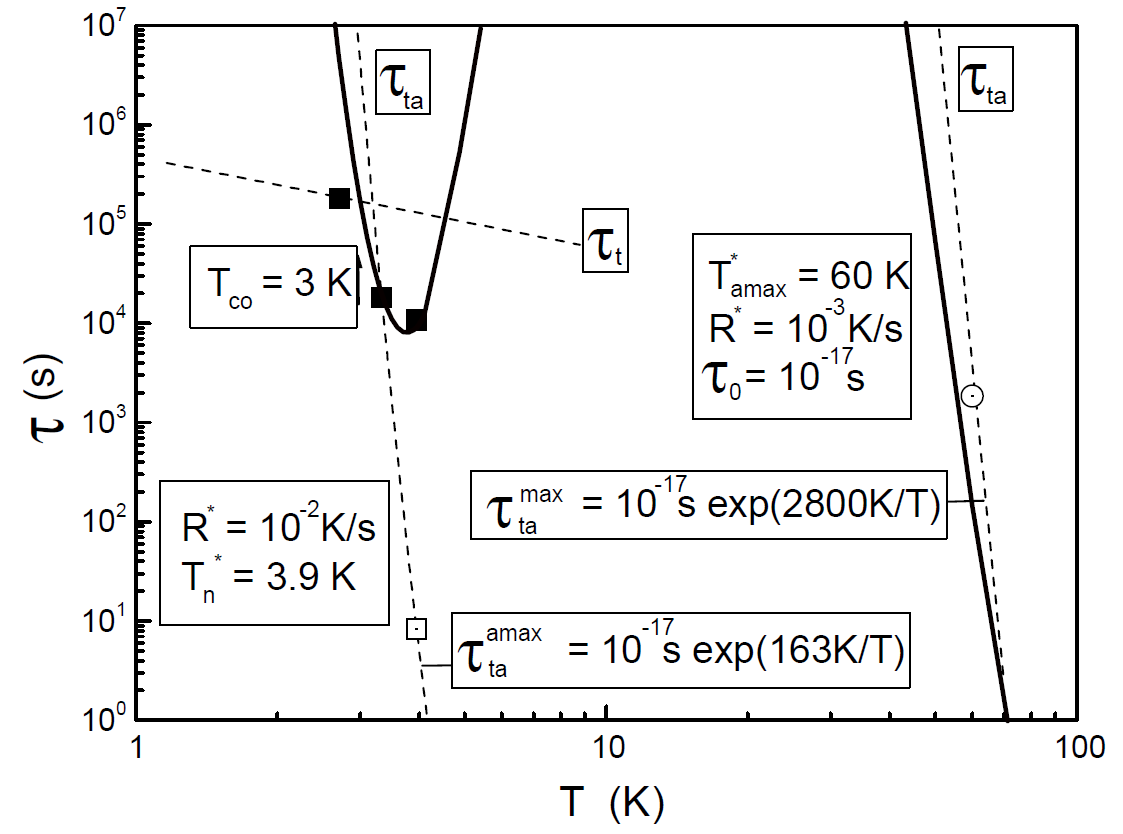}
\caption{The maximum relaxation time  $\tau_a^{max}$  used in the fit curves in Fig. 10 with Eq. (\ref{eq.21}) as a function of temperature. The value obtained at  4 K (full squares) is 3 orders of magnitude longer than expected from the Arrhenius law (open squares).}
\label{fig:12}
\end{figure}

For a better understanding, Fig.~\ref{fig:12} shows these data together with the calculated temperature dependencies within the STM. At the crossover temperature $T_{c0} = 3$ K the tunneling relaxation rate becomes equal to the rate of thermal activation. Thus the value of $\tau_a^{max}$ at the lowest temperature lies in the tunneling range, while two others are in the range of thermal activation. The Arrhenius law $\tau_{ta}^{-1}=\tau_0^{-1}\exp (-V/k_B T)$ with $\tau_0 = 10^{-17}$ s yields effective barrier heights $V_m/k_B = 163$ K ($T_0 = 3.33$ K) of the TLSs causing at this time the heat. It is expected within the STM (where barrier heights are temperature independent) that the corresponding relaxation time decreases very rapidly with increasing temperature (see Fig.~\ref{fig:12}, dashed line). For normal tunneling systems this leads to the saturation of the heat release at the freezing temperature  $T^*_n$, which is about 20\% higher than $T_{co}$, i.e.  $T^*_n=3.6$ K. However, no saturation was found up to $22$ K (see Fig.~\ref{fig:4}). In addition, the measured $\tau_a^{max}$ at $3.95$ K is $3$ orders of magnitude longer than expected from the Arrhenius law (open square in Fig.~\ref{fig:12}). All these facts demonstrate that the effective barrier height increases at higher temperature as shown in Fig.~\ref{fig:13}. Both the barrier  height and the corresponding relaxation time in Fig.~\ref{fig:12} were calculated with the following fit function (the curve in Fig.~\ref{fig:13}):
\begin{figure}[t]
\centering
\includegraphics[width=12.5cm,  angle=0]{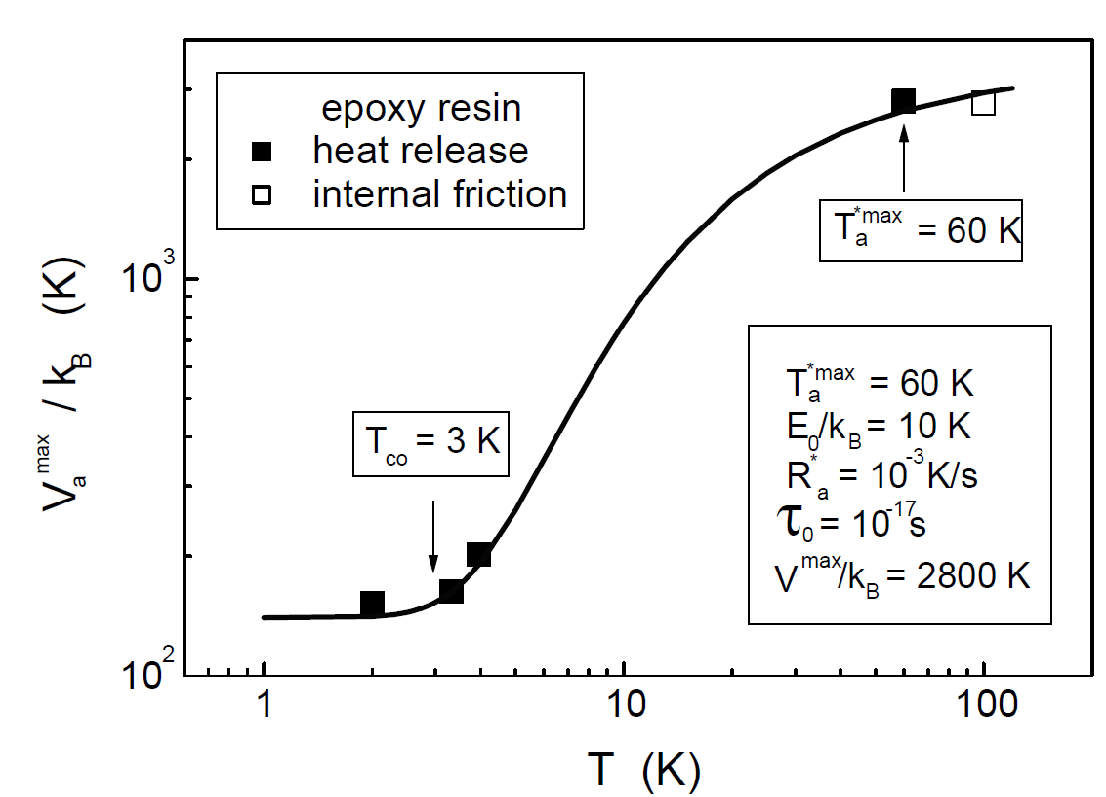}
\caption{The effective maximum barrier height  $V_a^{max}$ calculated from the heat release (full squares) and internal friction data  (open squares) of epoxy resin as a function of temperature. The curve corresponds to a fit with Eq. (\ref{eq.22}).}
\label{fig:13}
\end{figure}
\begin{equation}\label{eq.22}
 V^{max}_a/k_B = [140 + 3300 \exp(-16.5 \textrm{K}/T)] \textrm{K}.     
\end{equation}
We can estimate the freezing temperature  $T^*_a = 60$ K from the $T_1^2 - T_0^2$ dependence of the  parameter  $Q_1$ in Fig.~\ref{fig:4} by extrapolation of the experimental data at high and low $T_1$; finally we get from Eq. (8) in \cite{c1} the barrier height  $V_m/k_B = 2800$ K, which is close to  $V^{max}$.

\begin{figure}[t]
\centering
\includegraphics[width=12.5cm,  angle=0]{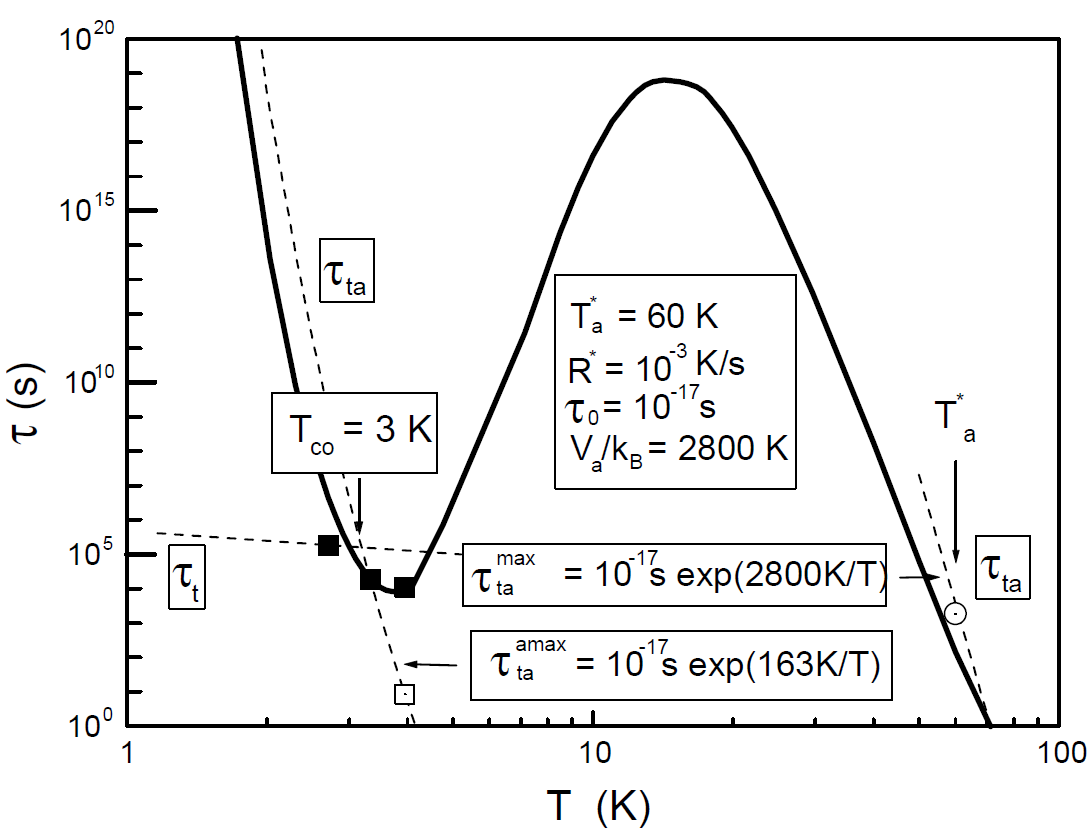}
\caption{The maximum relaxation time $\tau_a^{max}$ of epoxy resin as a function of temperature calculated with the temperature-dependent maximum barrier height $V_a^{max}$  in Fig. 12 (the curve). Full squares: $\tau_a^{max}$  deduced from the heat release data in Fig.  10, the open circle indicates the relaxation time at the freezing temperature deduced from the internal friction and heat release data.}
\label{fig:14}
\end{figure}

One can also estimate the maximum barrier height from the internal friction experiments. No maximum of the damping was observed in experiments with epoxy up to $100$ K ($f = 150$ MHz) \cite{c16}. This leads to a maximum barrier height larger than $1800$ K. A damping peak nearby $110$ K can be estimated for the TLSs in the vibrating reed experiments ($f = 84$ KHz) with stycast \cite{c17}. This yields  $V^{max}/k_B = 2860$ K (full squares in Fig.~\ref{fig:13}). All these parameters are in a reasonable relation. Thus, the TLSs with the maximum barrier heights become frozen in state at $60$ K (open circles in Figs.~\ref{fig:12} and \ref{fig:14}) and their relaxation time increases with decreasing temperature, reaches a maximum value of $10^{18}$ s at about $15$ K  than returns to a short time $10^4$ s at $4$ K (see Fig.~\ref{fig:14}). Let us note that the Arrhenius law gives $10^{303}$ s (!) for the relaxation time of TLSs with the maximum barrier height at $4$ K.  

\subsection{Heat release of pentanol-2}

Pentanol is a good candidate for a proton target due to its high hydrogen content (C$_5$H$_{11}$OH). However, a high polarization requires an uniform distribution of the paramagnetic centers, which is much better in amorphous  than crystalline materials. A standard cooling of pentanol with a dilution refrigerator yields polycrystalline pentanol. A difficult procedure was necessary to get the amorphous structure. The pentanol was mixed with $5$ \% water and cooled down rapidly by liquid nitrogen (the glass transition temperature $T_g$ lies near $170$ K). Then the solid pentanol was mounted rapidly in the dilution refrigerator and cooled down to lower temperatures. This procedure is not too convenient. Therefore we check Pentanol-2,  where the OH-group is transferred from the end of the long molecule to the side. Indeed, the worse symmetry of the molecules leads to an amorphous structure by normal cooling in a dilution refrigerator. Heat release experiments showed clearly a glassy behavior, however with deviations in the time dependence: the heat release was found to be proportional to  $t^{-0.69}$ between $20$ and $500$ min \cite{c18}. Eq.~(\ref{eq.17}) gives also a good fit of the time dependence. The corresponding  parameters  $Q_1$ are shown in Fig.~\ref{fig:4}. The crossover temperature is near $3$ K and one expects a saturation within the STM above $4$ K. However, the $T_1$-dependence does not saturate up to $12$ K. The data are close to that of epoxy resin (see Fig.~\ref{fig:4}).

\section{Discussion}

We can make three main conclusions from the above-reviewed experimental data for amorphous organic materials:
\begin{enumerate}
\item The main contribution to the heat release comes from anomalous TLSs. Indeed, the heat release saturates at very high starting temperatures $T_1$ (a giant heat release), the distribution parameter  $P_Q$ is close to  $P_C$ (or larger) and the heat release relaxes faster at very long time.
\item The heat release experiments with organic glasses correspond to the considered cases (1) and (2), where $t  \leq \tau_a^{max}$. In fact, $\tau_a^{max}$  in organic glasses is longer than in inorganic ones or glasslike crystalline materials, except NbTi/D (see Fig.~\ref{fig:3}).
\item  In contrast to inorganic materials, for a long time, but $t <  \tau_a^{max}$ the heat release is roughly proportional  to $t^{-a}$  with  $0.5 < a < 0.8$. 
\end{enumerate}
We also put emphasis on an abrupt decrease of the heat release with time in  3-mp/2,3-dmb, which cannot be described even by suggesting a cut in $P(V)$. 

Let us analyse these conclusions with relation to the model of local mechanical deformations due to giant large-scale fluctuations in thermal expansion during the cooling of a sample \cite{Karpov}. According to this model, the thermal expansion coefficients $\alpha(\bf r)$ take random values inside the dilatation centers of the sample. The dispersion in the distribution of $\alpha(\bf r)$ is suggested
to be much bigger than its mean value:  $\langle \alpha^2 \rangle/\langle \alpha \rangle^2\sim [D/(\Gamma E_{av})]^2\sim 10^5$, where   $D$ is the deformation potential, $\Gamma$ is the Gr\"{u}neisen parameter. Rapid cooling from an initial temperature $T_1$ to a final $T_0$ results in a generation of strong local mechanical stresses, which provokes a linear reduction in the barrier height. As a consequence, the life time of high-energy TLSs reduces and their heat release becomes experimentally observable. Namely these TLSs are called anomalous. 

This model allows us to explain qualitatively an increase of $\tau_a^{max}$ with an average energy of TLSs $E_{av}$ (see Fig.~\ref{fig:3}). Indeed, the dispersion $\langle\alpha^2\rangle$ decreases when $T_0+T_1$ grows. This results in growing barrier heights of anomalous TLSs, reducing of $\Delta_0$, and increasing relaxation time. Regarding the second item of our conclusions, the edge in the distribution function of anomalous TLSs is shifted to the left against the edge of normal TLSs on the value $\Delta V = \mp \sigma V_{ac}$. This step can be explained  under the assumption of a fixed limit of local stresses $\sigma^{max}$. Taking into account the structural characteristics of the materials one can suggest that $\sigma^{max}$ has a maximal value for inorganic crystalls, a little lesser for amorphous materials and has a smaller value for organic polymers. Indeed, the corresponding inverse behavior of $\tau_a^{max}$ takes place (see Fig.~\ref{fig:3}).

Deviations from the $t^{-1}$ behavior in the heat release can be related to the increase of the spectral density $P_0(t)\sim t^{1-a}$ during the cooling process. This means a gradual increase of local stresses and/or sizes of stress concentrated regions. In the last case, an increase of $P_0$ will be accompanied by a shift of $V^{max}$ to the left. This can explain an abrupt decrease of $\dot{Q}(t)$ in 3-mp/2,3-dmb. 

It should be stressed that there is no evidence of any deviations from $t^{-1}$ behavior in the heat release in inorganic materials. Therefore, a suggested increase of local stresses should be a specific property of organic materials. We suppose that local stresses are influenced by an expansion of the sample during the heat release process. Indeed, an average thermal expansion coefficient $\langle\alpha\rangle$  is related to $C_p$, Gr\"{u}neisen parameter  $\Gamma$ and isothermal compressibility $\kappa_T$ as 
\begin{equation}\label{Alp}
\langle\alpha\rangle = \kappa_T \Gamma C_p.
\end{equation}
According to (\ref{CP}), the heat capacity grows with time. At the same time, in organic polymers $\kappa_T$ and $\tau^{min}$ have a several orders bigger and
lesser values, respectively, in comparison to the inorganic materials. A decrease of $\tau^{min}$ is a result of strong electron-phonon interactions in polymers.  Hence in polymers we can expect a more pronounced influence of  $\langle\alpha\rangle$. Even a slow drift of $\langle\alpha\rangle$ can provoke a formation of additional local stresses  and, as a consequence, a raising quantity of anomalous TLSs. 

As another possible explanation of this phenomenon (which would also approve  Eq. (\ref{eq.17})) one can suggest the existence of an unknown additional channel which provides the relaxation of TLSs in the time $\tau_a^{max}$  independently of the barrier heights.  Unfortunately, the origin of this sort of  TLS relaxation is still an open question. A possible candidate is the resonant relaxation based on TLS-TLS interaction at very low temperatures suggested in ~\cite{Burin}.


\section{Summary}
Summarising, our analysis shows that all glasses (amorphous organic, inorganic, and glasslike crystalline materials) reveal an universal behavior of the relation between the heat release and the heat capacity in good agreement with our model of anomalous TLSs (see  Table \ref{tab1}).
The heat release shows a similar behavior in various organic polymers and has some peculiar properties in comparison with inorganic glasses and glasslike crystals. 
We have shown that despite these peculiarities all existing heat release data can be explained within the model of local mechanical stresses where specific properties of organic polymers are caused by a lower value of the maximum stress and a more pronounced drift of an average thermal expansion coefficient due to a bigger value of isothermal compressibility and stronger electron-phonon interaction. It should be stressed that organic glasses are convenient for further experimental investigations of the giant heat release owing to their longer $\tau_a^{max}$.
\begin{table}[h!]
\caption{Distribution parameters extracted from the heat release and the heat capacity measurements for different materials.}
\begin{center}
\begin{tabular}{|c|c|c|c|c|c|c|c|c|c|}
\hline
Material&$P_C$&$P_Q$&$P_C/P_Q$&$T_{co}$&$T^*_n$&$T^*_{a0}$& $T^*_a$ & case &Ref.\\
~& $10^{44}$ J$^{-1}$ m$^{-3}$& $10^{44}$ J$^{-1}$ m$^{-3}$& K&K&K&K&~&~\\
\hline
a-SiO$_2$& 8.0 &2.0&4.0&4.0&4.8&14.5&~&$t > \tau_a^{max}$&\cite{c5}\\
NbTi& 43.2 &5.3&8.2&4.4&5.4&47&~&$t > \tau_a^{max}$&\cite{c1}\\
NbTi$_{9\% \mathrm{H}}$&120&45&2.7&4.8&5.9&32&~&$t > \tau_a^{max}$&\cite{c1}\\
ZrO$_2$CaO&18.5&5.7&3.2&5.5&6.9&35&~&$t > \tau_a^{max}$&\cite{c7}\\
ZrO$_2$CaO&18.5&19.2&0.96&5.5&6.9&~&$>$ 64&$t < \tau_a^{max}$&\cite{c7}\\
NbTi$_{9\% \mathrm{D}}$&86&84&1.02&2.5&3.1&~&51&$t < \tau_a^{max}$&\cite{c1}\\
PMMA&4.8&28&0.30&$>$ 2&~&~&~&$t < \tau_a^{max}$&\cite{c9,c10}\\
Epoxy resin&10&7.8&1.28&2.0&2.6&~&60&$t < \tau_a^{max}$&\cite{c14}\\
PS&7.9&9.5&0.83&$>$ 1&~&~&~&$t < \tau_a^{max}$&\cite{c13}\\
3pm/2,3dmb&~&16.7&~&2.5&3.0&~&$>$20&$t < \tau_a^{max}$&\cite{c12}\\
pentanol&~&4.1&~&4.0&3.5&~&$>$12&$t < \tau_a^{max}$&\cite{c18}\\
\hline
\end{tabular}\label{tab1}
\end{center}
\end{table}
\newpage

This work has been supported by the
Heisenberg-Landau Program under Grant No. HLP-2013-26.
\newpage

\textbf{List of Symbols}
\begin{longtable}[H]{lp{0.9\linewidth}}
$\alpha$ & The thermal expansion coefficient.\\
$\Gamma$ & Gr\"{u}neisen parameter.\\
$\gamma_{l,t}$ & $\simeq \partial \Delta / 2 \partial u_{ik}$: Effective deformation potential for longitudinal or transversal phonons. \\
$\Delta_0$ & The tunneling energy.\\
$\kappa_T$ & Isothermal compressibility.\\
$\lambda$ &  The Gamow parameter.\\
$\rho$ & Mass density.\\
$\sigma$ & Mechanical stress.\\
$\tau_0$ & Pre-exponential factor in the Arrhenius law. \\
$\tau_a$ &  Relaxation time of tunneling systems due to phonon-assisted interaction for anomalous TLSs.\\
$\tau_t$ & Relaxation time of tunneling systems due to phonon-assisted interaction.\\
$\tau_{ta}$ & Thermally activated relaxation time of TLSs.\\
$\tau^{min}$ & Minimum relaxation time of TLSs.\\
$A$ & A parameter proportional to $\gamma^2/\rho \upsilon^5 \hbar^4$ with dimensions J$^{-3}$ s$^{-1}$\\
$C_p$ & The specific heat at constant pressure.\\
$D$ & Deformation potential of TLSs.\\
$E_0$ & The zero-point energy.\\
$E_{av}$ & Average energy of TLSs causing the heat release.\\
$m$ & The mass.\\
$P_0$ & Constant density of states of TLSs\\
$P_a$ &  Constant density of states of anomalous TLSs.\\
$P_C$ &  Constant density of states deduced from the heat capacity.\\
$P_n$ &  Constant density of states of normal TLSs.\\
$P_Q$ &  Constant density of states deduced from the heat release.\\
$P_{a0}$ &  Constant density of states of anomalous TLSs deduced from the heat release long time measurement ($t>\tau_a^{max}$).\\
$P_{ax}$ & A fit parameter of Eqs.~(\ref{eq.17}) and (\ref{eq.19}).\\
$R^*$ & Cooling rate in heat release experiments.\\
$T^*$ & Freezing temperature; below it and for typical cooling rates the TLSs remain in a  non-equilibrium state and contribute to the heat release.\\
$T_0$ & Measuring temperature in heat release experiments.\\
$T_1$ & ``Charging'' temperature in heat release experiments.\\
$T_{co}$ & Crossover temperature where the thermally activated relaxation time equals the tunneling relaxation time.\\
$T^*_a$ & Freezing temperature for anomalous TLSs.\\
$T^*_n$ & Freezing temperature for normal TLSs.\\
$Q_1$ & A fit parameter of Eq.~(\ref{eq.19}).\\
$Q_a$ & A fit parameter of Eq.~(\ref{eq.17}).\\
$Q_l$ & A fit parameter of Eq.~(\ref{eq.13})\\
$Q_n$ & A fit parameter of Eq.~(\ref{eq.17}).\\
$Q_s$ & A fit parameter of Eq.~(\ref{eq.13}).\\
$V$ & Potential barrier height. \\
$V_{a}$ & Potential barrier height of anomalous TLSs.\\
$V_m$ & The average barrier height of the TLSs causing the heat release (is directly proportional to the freezing temperature).\\
$V_s$ & Volume of the sample\\
$V_{ac}$ & Activation volume. \\
$\upsilon_{l,t}$ & The sound velocity.\\
\end{longtable}

\end{document}